\documentclass[12pt]{article}

\author{Yu.~M.~Zinoviev
       \thanks{E-mail address: Yurii.Zinoviev@ihep.ru} \\
        {\it Institute for High Energy Physics} \\
        {\it Protvino, Moscow Region, 142280, Russia}}
\title{On massive gravity and bigravity in three dimensions}

\date{}

\def\eptwo{\left\{ \phantom{|}^{\mu\nu}_{ab} \right\}}
\def\epthree{\left\{ \phantom{|}^{\mu\nu\alpha}_{abc} \right\}}

\begin{document}

\maketitle

\begin{abstract}
In this paper we investigate possible consistent ghost-free models
containing massive spin 2 particles in three dimensions. We work in a
constructive approach based on the frame-like gauge invariant
description for such massive spin 2 particles. We provide the most
general form of linear approximations, i.e. cubic vertices in the
Lagrangian and linear in fields corrections to gauge transformations.
As for the possibility to go beyond the linear approximation, we show
that  there  exists at least one solution that admits non-singular
massless limit and that corresponds to a so called "New massive
gravity".
\end{abstract}

\thispagestyle{empty}
\newpage
\setcounter{page}{1}

\section*{Introduction}

Constructing consistent interacting theories containing massive spin 2
particles is an old, interesting and important physical problem. One
of the main difficulties one faces in such theories is the appearance
of non-physical ghost degree of freedom \cite{BD72}. During last
three years essential progress has been achieved in this direction.

In three dimensions a so called "New massive gravity" appeared
\cite{BHT09,BHT09a}. It was constructed as a particular example of
higher derivatives gravity but it turns out to be equivalent to the
system of massless and massive spin 2 particles, the massless one
being a ghost. To a great extent such construction is specific namely
to spin 2 in three dimensions so that it is not an easy and
straightforward task to find its generalizations to higher spins (see
e.g. \cite{BHT09b,BKRTY11}) or higher dimensions
(e.g. \cite{NO09,GT09,LP11,BFRT12}). One of the open questions is the
so called partially massless limit which exists for the free massive
spin 2 in de Sitter space and where additional local gauge symmetry
arises \cite{BC11}.

More recently a whole family of consistent ghost-free models in four
dimensions has been constructed both the massive gravity
\cite{RG10,RGT10}, as well as for massive bigravity
\cite{HRS11,HR11}.
In general such model consist of usual action for one or two massless
gravitons (non interacting in the massless limit) and complicated
non-linear potential terms without derivatives. In this, there is no
any particular symmetry that can guarantee and/or explain the absence
of ghost degree of freedom, so to check that one has to go through
careful Hamiltonian analysis \cite{HR11a,HR11b}. Even in the so
called Stueckelberg formulation \cite{RGT11} where gauge symmetries of
massless theory are restored, to check the absence of ghost one still
have to use Hamiltonian analyses \cite{HSS12}. As in the three
dimensional case, it is not at all clear what happens in such theories
in the partially massless limit.

In both cases it seems that it would be instructive if we can
reproduce such theories in a constructive approach based on the gauge
invariant description for massive higher spin particles
\cite{Zin08b,PV10}. Such formalism has enough gauge symmetries to
guarantee (and explain) the absence of ghosts without using careful
Hamiltonian analysis. Also it seems natural to work in a frame-like
formalism where the structure of potential terms becomes much more
simple and clear \cite{HR12}. In this paper we begin such a program
starting with the $d=3$ case.

The plan of the paper is simple. Section 1 devoted to the massless
case. First of all we briefly remind a frame-like description of
$d=3$ massless gravity (just to set notations and conventions) and
then consider the most general interacting theory for two massless
ones. Main Section 2 devoted to the case where one of spin 2 particles
is massive, while the other one remains massless. In Subsection 2.1 we
give frame-like gauge invariant description of massive spin 2
particles \cite{Zin08b} adopted to $d=3$ case. The in Subsections 2.2
and 2.3 we consider in linear approximation self-interaction for
massive spin 2 and its interaction with massless graviton,
respectively. At last, in Subsection 2.4 we discuss possibilities to
go beyond linear approximation. we show that if we are looking for the
theory admitting non-singular massless limit that reduce to massless
bi-gravity considered in Section 1, then there exists at least one
solution which requires that massless graviton be a ghost exactly as
in New massive gravity.

\section{Massless case}

In this section we consider massless gravity and bigravity as
starting point for models where one of the spin 2 particles becomes
massive while the other one remains massless.

\subsection{Gravity}

Usually frame-like formalism for gravity involves pair of fields ---
frame $h_\mu{}^a$ and Lorentz connection $\omega_\mu{}^{ab} = -
\omega_\mu{}^{ba}$. But in three dimensions it is very convenient to
use dual variable $\omega_\mu{}^{ab} \to \omega_\mu{}^a =
\varepsilon^{abc} \omega_\mu{}^{bc}$. In these notations the free
Lagrangian describing massless spin 2 particles in $(A)dS_3$ space can
be written as follows (parameter $\sigma = \pm 1$ takes into account
that in $d=3$ massless spin 2 may be a ghost):
\begin{equation}
\sigma {\cal L}_0 = \frac{1}{2} \eptwo \omega_\mu{}^a \omega_\mu{}^b -
\varepsilon^{\mu\nu\alpha} \omega_\mu{}^a D_\nu h_\alpha{}^a -
\frac{\Lambda}{2} \eptwo h_\mu{}^a h_\nu{}^b
\end{equation}
Here $\Lambda$ --- cosmological constant, $\eptwo = e^\mu{}_a
e^\nu{}_b - e^\mu{}_b e^\nu{}_a$ and so on, where $e_\mu{}^a$ ---
non-dynamical background frame while $AdS_3$ covariant derivatives
$D_\mu$ are normalized so that
$$
[ D_\mu, D_\nu ] \xi^a = - \Lambda e_{[\mu}{}^a \xi_{\nu]}
$$
This Lagrangian is invariant under the following local gauge
transformations:
\begin{equation}
\delta_0 h_\mu{}^a = D_\mu \hat{\xi}^a + \varepsilon_\mu{}^{ab}
\hat{\eta}^b, \qquad \delta_0 \omega_\mu{}^a = D_\mu \hat{\eta}^a -
\Lambda \varepsilon_\mu{}^{ab} \hat{\xi}^b
\end{equation}
where $\hat{\eta}^a$ --- dual to Lorentz transformation parameter
$\hat{\eta}^a = \varepsilon^{abc} \hat{\eta}^{bc}$.

In a frame-like formalism it is easy to introduce self-interaction for
such massless spin 2 particles in the linear
approximation\footnote{Here and in what follows linear approximation
means cubic vertices in the Lagrangian and linear in fields
corrections to gauge transformations, hence the name.}. 
Cubic vertex has the form
\begin{equation}
{\cal L}_1 = \kappa_0 \epthree [ h_\mu{}^a \omega_\nu{}^b
\omega_\alpha{}^c - \frac{\Lambda}{3} h_\mu{}^a h_\nu{}^b
h_\alpha{}^c ] 
\end{equation}
where $\kappa_0$ --- coupling constant while corresponding corrections
to gauge transformations look as follows:
\begin{eqnarray}
\delta_1 h_\mu{}^a &=& - 2\sigma\kappa_0 \varepsilon^{abc} [ h_\mu{}^b
\hat{\eta}^c + \omega_\mu{}^b \hat{\xi}^c] \nonumber \\
\delta_1 \omega_\mu{}^a &=& - 2\sigma\kappa_0 \varepsilon^{abc} [
\omega_\mu{}^b \hat{\eta}^c - \Lambda h_\mu{}^b \hat{\xi}^c ]
\end{eqnarray}
A remarkable feature of $d=3$ frame-like formalism is that there are
no any quartic vertices for spin 2 (and all spins higher than 2)
particles. Thus for the theory to be closed we must have $\delta_1
{\cal L}_1 = 0$. For the case at hands it is easy to check that these
variations indeed cancel.

\subsection{Bigravity}

Recall that in general $d \ge 3$ dimension there are only two possible
cases for interacting theories with two massless spin 2 particles
(see e.g. \cite{Wald86,Wald86a,BDGH00}). In the first one, where both
spin 2 particles are physical, any interacting Lagrangian by field
redefinitions can be reduced to the sum of two independent halves. In
the second one we do have non-trivial cross-interaction with the price
that one of the spin 2 particles must be a ghost so that such case is
of interest in $d=3$ only. Let us see how these results come in $d=3$
frame-like formalism.

We will use the following notations for second spin 2 particle and its
gauge parameters: $\Omega_\mu{}^a$, $f_\mu{}^a$, $\eta^a$
and $\xi^a$. Let us consider interactions in the linear approximation.
There are four possible cubic vertices which we denote $hhh$, $hhf$,
$hff$ and $fff$ correspondingly. In the linear approximation they are
completely independent from each other so we can consider them
separately.\\
{\bf Vertex $hhh$} is the same as in the previous subsection.\\
{\bf Vertex $hhf$} Here cubic vertex has the form
\begin{equation}
{\cal L}_1 = \kappa_1 \epthree [ f_\mu{}^a \omega_\nu{}^b
\omega_\alpha{}^c + 2 h_\mu{}^a \omega_\mu{}^b \Omega_\alpha{}^c -
\frac{\Lambda}{2} f_\mu{}^a h_\nu{}^b h_\alpha{}^c ]
\end{equation}
while corrections to gauge transformations look like:
\begin{eqnarray}
\delta_1 \omega_\mu{}^a &=& - 2\kappa_1 \varepsilon^{abc} [
\Omega_\mu{}^b \hat{\eta}^c + \omega_\mu{}^b \eta^c - \Lambda
f_\mu{}^b \hat{\xi}^c - \Lambda h_\mu{}^b \xi^c ] \nonumber \\
\delta_1 h_\mu{}^a &=& - 2\kappa_1 \varepsilon^{abc} [ f_\mu{}^b
\hat{\eta}^c + h_\mu{}^b \eta^c + \Omega_\mu{}^b \hat{\xi}^c +
\omega_\mu{}^b \xi^c ] \nonumber \\
\delta_1 \Omega_\mu{}^a &=& - 2\kappa_1 \varepsilon^{abc} [
\omega_\mu{}^b \hat{\eta}^c - \Lambda h_\mu{}^b \hat{\xi}^c ]
\label{ckap1} \\
\delta_1 f_\mu{}^a &=& - 2\kappa_1 \varepsilon^{abc} [ h_\mu{}^b
\hat{\eta}^c + \omega_\mu{}^b \hat{\xi}^c ] \nonumber
\end{eqnarray}
{\bf Vertex $hff$} This case is similar to the previous one but roles
of two fields are interchanged:
\begin{equation}
{\cal L}_1 = \kappa_2 \epthree [ h_\mu{}^a \Omega_\nu{}^b
\Omega_\alpha{}^c + 2 f_\mu{}^a \omega_\nu{}^b \Omega_\alpha{}^c -
\frac{\Lambda}{2} h_\mu{}^a f_\nu{}^b f_\alpha{}^c ]
\end{equation}
\begin{eqnarray}
\delta_1 \omega_\mu{}^a &=& - 2\kappa_2 \varepsilon^{abc} [
\Omega_\mu{}^b \eta^c - \Lambda f_\mu{}^b \xi^c ] \nonumber \\
\delta_1 h_\mu{}^a &=& - 2\kappa_2 \varepsilon^{abc} [ f_\mu{}^b
\eta^c + \Omega_\mu{}^b \xi^c ] \\
\delta_1 \Omega_\mu{}^a &=& - 2\kappa_2 \varepsilon^{abc} [
\Omega_\mu{}^b \hat{\eta}^c + \omega_\mu{}^b \eta^c - \Lambda
f_\mu{}^b \hat{\xi}^c - \Lambda h_\mu{}^b \xi^c ] \nonumber \\
\delta_1 f_\mu{}^a &=& - 2\kappa_2 \varepsilon^{abc} [ f_\mu{}^b
\hat{\eta}^c + h_\mu{}^b \eta^c + \Omega_\mu{}^b \hat{\xi}^c +
\omega_\mu{}^b \eta^c ] \nonumber
\end{eqnarray}
{\bf Vertex $fff$} And this case is similar to $hhh$ one:

\begin{equation}
{\cal L}_1 = \kappa_3 \epthree [ f_\mu{}^a \Omega_\nu{}^b
\Omega_\alpha{}^c - \frac{\Lambda}{3} f_\mu{}^a f_\nu{}^b f_\alpha{}^c
]
\end{equation}
\begin{eqnarray}
\delta_1 \Omega_\mu{}^a &=& - 2\kappa_3 \varepsilon^{abc} [
\Omega_\mu{}^b \eta^c - \Lambda f_\mu{}^b \xi^c ] \nonumber \\
\delta_1 f_\mu{}^a &=& - 2\kappa_3 \varepsilon^{abc} [ f_\mu{}^b
\eta^c + \Omega_\mu{}^b \xi^c ]
\end{eqnarray}

Note that as can be seen from formulas given above gauge
transformations mix two our spin 2 fields so that if we try to go
beyond linear approximation these four cubic vertices will not be
independent any more. And here we again face the fact that in $d=3$
frame-like formalism there are no any quartic vertices so that all
variations of cubic ones must cancel each other. Happily, this is
indeed possible provided the following relation holds:
\begin{equation}
\kappa_1{}^2 + \kappa_2{}^2 - \sigma\kappa_0\kappa_2 -
\kappa_1\kappa_3 = 0 \label{rel}
\end{equation}
Thus we have solution with three parameters and their meaning is
rather clear: we have two spin 2 particles and thus two independent
coupling constants and also a kind of "mixing angle". This last
parameter is related with the fact that we have two similar particles
and so we can make field redefinition mixing them. But in the case
where one of the particles become massive while the other one remains
massless this symmetry between them is broken so we will not try to
make such redefinition\footnote{Clearly, we still can do such field
redefinition but as a result mass terms will not be diagonal any
more.}. Instead, we will use the fact that there are severe
restrictions on the possible cubic vertices with two massless and one
massive spin particles. In general $d \ge 4$ case \cite{Met05,Met12}
such vertex requires as many as 6 derivatives and in $d=3$ it is just
absent (see Appendix B). Thus, assuming that it is the second particle
($f_\mu{}^a$, $\Omega_\mu{}^a$) that will become massive, we must set
$\kappa_1 = 0$, in this from the relation (\ref{rel}) we immediately
obtain that\footnote{This relation is nothing but usual manifestation
of universality of gravitational interactions, i.e. the same coupling
constant determines both self-interaction for graviton as well as its
interaction with matter with massive spin 2 playing the role of matter
here.} $\kappa_2 = \sigma\kappa_0$ while $\kappa_3$ remains arbitrary.

\section{Massive case}

In this section we consider models combining massless spin 2 particle
(that may be physical one or ghost) and massive one. We will work in a
constructive approach using frame-like gauge invariant description for
massive spin 2 particles. General $d \ge 3$ case has been constructed
in \cite{Zin08b} (see also \cite{PV10}) and here we give version
adopted to $d = 3$ dimensions.

\subsection{Gauge invariant frame-like formalism}

For the description of massive spin 2 particle in $(A)dS_3$ we will
use the following set of fields: ($\Omega_\mu{}^a$, $f_\mu{}^a$),
($B^a$, $A_\mu$) and ($\pi^a$, $\varphi$), where $B^a =
\varepsilon^{abc} F^{bc}$. Then the free Lagrangian has the form:
\begin{eqnarray}
{\cal L}_0 &=& \frac{1}{2} \eptwo \Omega_\mu{}^a \Omega_\nu{}^b -
\varepsilon^{\mu\nu\alpha} \Omega_\mu{}^a D_\nu f_\alpha{}^a +
\frac{1}{2} B_a{}^2 - \varepsilon^{\mu\nu\alpha} B_\mu D_\nu
A_\alpha - \frac{1}{2} \pi_a{}^2 + \pi^\mu D_\mu \varphi + \nonumber
\\ 
 && + m \varepsilon^{\mu\nu\alpha} [ -2 \Omega_{\mu\nu} A_\alpha +
B_\mu f_{\nu\alpha} ] + 2M  \pi^\mu A_\mu + \nonumber \\
 && + \frac{M^2}{2} \eptwo f_\mu{}^a f_\nu{}^b + 2mM e^\mu{}_a
f_\mu{}^a \varphi + 3m^2 \varphi^2
\end{eqnarray}
where $M^2 = 2m^2 - \Lambda$. This Lagrangian is invariant under the
following local gauge transformations:
\begin{eqnarray}
\delta_0 \Omega_\mu{}^a &=& D_\mu \eta^a + M^2 \varepsilon_\mu{}^{ab}
\xi^b \nonumber \\
\delta_0 f_\mu{}^a &=& D_\mu \xi^a + \varepsilon_\mu{}^{ab}
\eta^b + 2m e_\mu{}^a \xi \\ 
\delta_0 B^a &=& - 2m \eta^a, \qquad
\delta_0 A_\mu = D_\mu \xi + m \xi_\mu \nonumber \\
\delta_0 \pi^a &=& 2mM \xi^a, \qquad
\delta_0 \varphi = - 2M  \xi \nonumber
\end{eqnarray}

Recall that in $dS$ space ($\Lambda > 0$) there exists a so called
partially massless limit $M \to 0$, where scalar field completely
decouples, leaving us with the Lagrangian
\begin{eqnarray}
{\cal L}_0 &=& \frac{1}{2} \eptwo \Omega_\mu{}^a \Omega_\nu{}^b -
\varepsilon^{\mu\nu\alpha} \Omega_\mu{}^a D_\nu f_\alpha{}^a +
\frac{1}{2} B_a{}^2 - \varepsilon^{\mu\nu\alpha} B_\mu D_\nu
A_\alpha + \nonumber \\ 
 && + m \varepsilon^{\mu\nu\alpha} [ -2 \Omega_{\mu\nu} A_\alpha +
B_\mu f_{\nu\alpha} ] 
\end{eqnarray}
which is still invariant under all three gauge transformations
\begin{eqnarray}
\delta_0 \Omega_\mu{}^a &=& D_\mu \eta^a, \qquad
\delta_0 f_\mu{}^a = D_\mu \xi^a + \varepsilon_\mu{}^{ab}
\eta^b + 2m e_\mu{}^a \xi \nonumber \\ 
\delta_0 B^a &=& - 2m \eta^a, \qquad
\delta_0 A_\mu = D_\mu \xi + m \xi_\mu 
\end{eqnarray}
As a result we obtain system with only one physical degree of freedom
instead of two in general massive case.

\subsection{Self-interaction}

In this subsection we consider possible self-interaction for massive
spin 2 particle. As we have already mentioned we will work in a
constructive approach where one construct the most general terms for
the Lagrangian and corrections to gauge transformations and requires
that the whole Lagrangian will be gauge invariant. In massive case
due to large number of fields such calculations turn out to be rather
complicated thus it is important to group different variations in some
convenient way. In a metric-like formalism (see e.g. \cite{Zin06}) one
may use grouping by the number of derivatives, while in a frame-like
formalism where both physical and auxiliary fields are present it is
convenient to group them by the mass order of coefficients. Thus for
the free Lagrangian we will have ${\cal L}_0 = {\cal L}_{00} + 
{\cal L}_{01} + {\cal L}_{02}$ where ${\cal L}_{00}$ --- kinetic terms
while ${\cal L}_{01}$ and ${\cal L}_{02}$ contains terms of order $m$
and $m^2$ respectively. Similarly in the linear approximation we will
write cubic vertices and linear corrections to gauge transformations
as
\begin{equation}
{\cal L}_1 = {\cal L}_{10} + {\cal L}_{11} + {\cal L}_{12}, \qquad
\delta_1 = \delta_{10} + \delta_{11} + \delta_{12}
\end{equation}
This implies that we begin with some massless theory satisfying
$$
\delta_{00} {\cal L}_{10} + \delta_{10} {\cal L}_{00} = 0
$$
and then we proceed with the deformation of such theory to non-zero
mass considering variations of order $m$:
$$
\delta_{00} {\cal L}_{11} + \delta_{01} {\cal L}_{10} +
\delta_{10} {\cal L}_{01} + \delta_{11} {\cal L}_{00} = 0
$$
and so on.

Let us begin with ${\cal L}_{10}$. In Appendix A we show that all
possible terms containing two spin 2 and one spin 0 particles can be
removed by appropriate fields redefinitions. Taking into account the
absence of such $2-2-0$ vertex the most general form can be written as
follows:
\begin{eqnarray}
{\cal L}_{10} &=& \kappa_3 \epthree \Omega_\mu{}^a \Omega_\nu{}^b
f_\alpha{}^c + a_1 f B^a B^a + a_2 \varepsilon^{\mu\nu\alpha}
f_\mu{}^a B^a D_\nu A_\alpha + a_3 \varphi B^a B^a + \nonumber \\
 && + a_4 \varphi \varepsilon^{\mu\nu\alpha} B_\mu D_\nu A_\alpha +
a_5 f \pi^a \pi^a + a_6 \eptwo f_\mu{}^a D_\nu \varphi \pi^b
\end{eqnarray}
Note that possible terms of the form $\varphi \pi^2$ and $\varphi \pi
D \varphi$ can also be removed by field redefinitions
$$
\pi^a \Rightarrow \pi^a + \kappa_1 \varphi \pi^a, \qquad
\varphi \Rightarrow \varphi + \kappa_2 \varphi^2
$$
There is one more possible redefinition
\begin{equation}
B^a \Rightarrow B^a + \kappa_0 \varphi B^a \label{redef}
\end{equation}
that we will use later on. Let us consider variations of order
$m^0$.\\
{\bf $\eta^a$ transformations:}
\begin{eqnarray*}
&& 2\kappa_3 \epthree [ D_\mu \Omega_\nu{}^a \eta^b f_\alpha{}^c -
\Omega_\mu{}^a \eta^b D_\nu f_\alpha{}^c ] + 2a_0
\varepsilon^{\mu\nu\alpha} \Omega_{\mu,\nu} \Omega_\alpha{}^a \eta^a +
\\
&& + a_2 \eptwo B^a \eta^b D_\mu A_\nu - a_6
\varepsilon^{\mu\nu\alpha} \pi_\mu D_\nu \varphi \eta_\alpha
\end{eqnarray*}
To compensate these variations we introduce the following corrections
to gauge transformations\footnote{In a frame-like formalism the
structure of such corrections is completely determined by the terms in
variations containing explicit derivatives.}:
\begin{eqnarray*}
\delta f_\mu{}^a &=& - 2\kappa_3 \varepsilon^{abc} f_\mu{}^b \eta^c,
\qquad \delta \Omega_\mu{}^a = - 2\kappa_3 \varepsilon^{abc}
\Omega_\mu{}^b \eta^c \\
\delta B^a &=& \alpha_1 \varepsilon^{abc} B^b \eta^c,
\qquad \delta \pi^a = \alpha_2 \varepsilon^{abc} \pi^b \eta^c
\end{eqnarray*}
This gives $a_2 = \alpha_1$ and $a_6 = - \alpha_2$.\\
{\bf $\xi^a$ transformations:}
\begin{eqnarray*}
&& - 2\kappa_3 \epthree D_\mu \Omega_\nu{}^a \Omega_\alpha{}^b \xi^c -
2a_1 \xi^\mu B^a D_\mu B^a - a_2 \varepsilon^{\mu\nu\alpha} \xi^a
D_\mu B^a D_\nu A_\alpha - \\
&& - 2a_5 \xi^\mu \pi^a D_\mu \pi^a + a_6 \eptwo D_\mu \pi^a D_\nu
\varphi \xi^b
\end{eqnarray*}
Thus we need the following corrections:
\begin{eqnarray*}
\delta f_\mu{}^a &=& - 2\kappa_3 \varepsilon^{abc} \Omega_\mu{}^b
\xi^c, \qquad \delta A_\mu = \alpha_3 \varepsilon_\mu{}^{ab} B^a
\xi^b, \qquad \delta B_\mu = \alpha_4 \xi^a D_\mu B^a \\
\delta \varphi &=& \alpha_5 (\pi \xi), \qquad
\delta \pi^a = \alpha_6 ( \xi^\mu D_\mu \pi^a - \xi^a (D\pi))
\end{eqnarray*}
In this, all variations can be cancelled provided
$$
2a_1 = \alpha_3 = \alpha_4 = - \alpha_1, \qquad
2a_5 = - \alpha_5 = - \alpha_6 = \alpha_2
$$
Thus in this order we obtain:
\begin{eqnarray}
{\cal L}_{10} &=&  \kappa_3 \epthree \Omega_\mu{}^a \Omega_\nu{}^b
f_\alpha{}^c - \frac{\alpha_1}{2} f B^a B^a + \alpha_1
\varepsilon^{\mu\nu\alpha} f_\mu{}^a B^a D_\nu A_\alpha + \nonumber \\
 && + a_3 \varphi B^a B^a + a_4 \varphi \varepsilon^{\mu\nu\alpha}
B_\mu D_\nu A_\alpha + \frac{\alpha_2}{2} f \pi^a \pi^a - \alpha_2
\eptwo f_\mu{}^a D_\nu \varphi \pi^b \label{l10}
\end{eqnarray}
\begin{eqnarray}
\delta_{10} \Omega_\mu{}^a &=& - 2\kappa_3 \varepsilon^{abc}
\Omega_\mu{}^b \eta^c, \qquad \delta_{10} f_\mu{}^a = - 2\kappa_3
\varepsilon^{abc} [ f_\mu{}^b \eta^c + \Omega_\mu{}^b \xi^c ]
\nonumber \\
\delta_{10} B_\mu &=& \alpha_1 \varepsilon_\mu{}^{ab} B^a \eta^b -
\alpha_1 \xi^a D_\mu B^a, \qquad \delta_{10} A_\mu = - \alpha_1
\varepsilon_\mu{}^{ab} B^a \xi^b \label{c10} \\
\delta_{10} \pi^a &=& \alpha_2 \varepsilon^{abc} \pi^b \eta^c -
\alpha_2 (\xi^\mu D_\mu \pi^a - \xi^a (D\pi)), \qquad \delta_{10}
\varphi = - \alpha_2 (\pi\xi) \nonumber
\end{eqnarray}
Let us consider variations of order $m$. The most general terms in the
Lagrangian have the form:
\begin{eqnarray*}
{\cal L}_{11} &=& \varepsilon^{\mu\nu\alpha} [ b_1 f_\mu{}^a
\Omega_\nu{}^a A_\alpha + b_2 f_{\mu,\nu} f_\alpha{}^a B^a + b_3
\Omega_{\mu,\nu} A_\alpha \varphi + b_4 f_{\mu,\nu} B_\alpha \varphi ]
\\
&& + b_5 \eptwo f_\mu{}^a A_\nu \pi^b + b_6 \varphi (\pi A)
\end{eqnarray*}
{\bf $\eta^a$ transformations:}
\begin{eqnarray*}
&& \varepsilon^{\mu\nu\alpha} [ b_1 D_\mu f_\nu{}^a \eta^a A_\alpha - 
(b_1+2m\alpha_1) f_\mu{}^a \eta^a D_\nu A_\alpha + (b_3-2ma_4)
\eta_\mu D_\nu A_\alpha \varphi - b_3 \eta_\mu A_\nu D_\alpha \varphi
] + \\
&& + (b_1-4m\kappa_3) \eptwo \Omega_\mu{}^a A_\nu \eta^b + 2b_2
\eta^\mu
f_\mu{}^a B^a + (b_2+m\alpha_1) \eptwo f_{\mu,\nu} B^a \eta^b + \\
&& + 2m\alpha_1 f (B \eta) + (2b_4-4ma_3) \varphi (B\eta) 
+ 2m\kappa_3 \eptwo f_\mu{}^a B_\nu \eta^b + (b_5+2M\alpha_2)
\varepsilon^{\mu\nu\alpha} \pi_\nu \eta_\alpha A_\mu
\end{eqnarray*}
The most general form of corrections would be:
$$
\delta \Omega_\mu{}^a = \beta_1 A_\mu \eta^a, \qquad
\delta B_\mu = \beta_2 f_\mu{}^a \eta^a + \beta_3 \varphi \eta_\mu
$$
but here we use remaining field redefinition (\ref{redef}) and put
$\beta_3 = 0$. Then all variations can be cancelled provided
$$
\alpha_1 = - \kappa_3, \qquad \beta_1 = 2m\kappa_3, \qquad
\beta_2 = 0, \qquad a_4 = 0
$$
$$
b_1 = 2m\kappa_3, \qquad b_2 = m\kappa_3, \qquad b_3 = 0,
\qquad b_4 = 2ma_3, \qquad b_5 = - 2M\alpha_2
$$
{\bf $\xi^a$ transformations:}
$$
\varepsilon^{\mu\nu\alpha} [ - b_1 D_\mu \Omega_\nu{}^a A_\alpha \xi^a
- b_2 D_\mu f_{\nu,\alpha} (B\xi) +  b_1 \Omega_\mu{}^a \xi^a D_\nu
A_\alpha  ] + \eptwo [ 2m\alpha_1 \Omega_{\mu,\nu} B^a \xi^b +
2m\kappa_3 \Omega_\mu{}^a B_\nu \xi^b ] 
$$
$$
(b_4+ma_4) \varepsilon^{\mu\nu\alpha} \xi_\mu D_\nu
B_\alpha \varphi - b_5 \eptwo \xi^a \pi^b D_\mu A_\nu 
- (b_4+ma_4) \varepsilon^{\mu\nu\alpha} \xi_\mu B_\nu D_\alpha \varphi
- 2M\alpha_1 \varepsilon^{\mu\nu\alpha} \pi_\mu B_\nu \xi_\alpha 
$$
Thus we need the following corrections:
\begin{eqnarray*}
\delta \Omega_\mu{}^a &=& - m\kappa_3 e_\mu{}^a (B\xi), \qquad
\delta f_\mu{}^a = - 2m\kappa_3 A_\mu \xi^a \\
\delta B_\mu &=& 2n\kappa_3 \Omega_\mu{}^a \xi^a - 2M\alpha_2
\varepsilon_\mu{}^{ab} \pi^a \xi^b, \qquad \delta A_\mu = 2ma_3
\varphi \xi_\mu \\
\delta \pi^a &=& - 2ma_3 \varepsilon^{abc} B^b \xi^c
\end{eqnarray*}
In this, cancellation of such variations requires
$$
2M(\alpha_2 - \alpha_1) + 2ma_3 = 0
$$
{\bf $\xi$ transformations:}
\begin{eqnarray*}
&& - 2m\kappa_3 \varepsilon^{\mu\nu\alpha} [ D_\mu f_\nu{}^a
\Omega_\alpha{}^a \xi - f_\mu{}^a D_\nu \Omega_\alpha{}^a \xi ] +
2m\kappa_3 \eptwo \Omega_\mu{}^a \Omega_\nu{}^b \xi - \\
&& - (3m\alpha_1+2Ma_3) B^a B^a \xi + (2m\alpha_1-2Ma_4)
\varepsilon^{\mu\nu\alpha} B_\mu D_\nu A_\alpha \xi - \\
&& - (b_6+4m\alpha_2) \pi^\mu D_\mu \varphi \xi - b_6 \varphi (D\pi)
\xi  + 3m\alpha_2 \pi^a \pi^a \xi 
\end{eqnarray*}
This time we introduce corrections of the form:
\begin{eqnarray*}
\delta \Omega_\mu{}^a &=& - 2m\kappa_3 \Omega_\mu{}^a \xi, \qquad
\delta f_\mu{}^a = 2m\kappa_3 f_\mu{}^a \xi \\
\delta B^a &=& (2m\alpha_1 - 2Ma_4) B^a \xi \\
\delta \pi^a &=& 3m\alpha_2 \pi^a \xi, \qquad \delta \varphi =
m\alpha_2 \varphi \xi
\end{eqnarray*}
In this, all such variations cancel provided
$$
2M(a_3 + a_4) = - m\alpha_1, \qquad b_6 = - m\alpha_2
$$
Note that combining the results from $\xi^a$ and $\xi$ transformations
we obtain:
\begin{equation}
a_3 = - \frac{m\alpha_1}{2M}, \qquad
\alpha_2 = (1 + \frac{m^2}{2M^2}) \alpha_1
\end{equation}
Collecting all pieces together we obtain:
\begin{eqnarray}
{\cal L}_{11} &=& m\varepsilon^{\mu\nu\alpha} [ 2\kappa_3 f_\mu{}^a
\Omega_\nu{}^a A_\alpha - \alpha_1 f_{\mu,\nu} f_\alpha{}^a B^a +
2a_3 f_{\mu,\nu} B_\alpha \varphi ] - \nonumber \\
 && - 2M\alpha_2 \eptwo f_\mu{}^a A_\nu \pi^b - m\alpha_2 \varphi 
(\pi A) \label{l11}
\end{eqnarray}
\begin{eqnarray}
\delta_{11} \Omega_\mu{}^a &=& 2m\kappa_3 A_\mu \eta^a - m\kappa_3
e_\mu{}^a (B\xi) - 2m\kappa_3 \Omega_\mu{}^a \xi, \qquad
\delta_{11} f_\mu{}^a = - 2m\kappa_3 A_\mu \xi^a + 2m\kappa_3
f_\mu{}^a \xi \nonumber \\
\delta_{11} B_\mu &=& 2m\kappa_3 \Omega_\mu{}^a \xi^a - 2M\alpha_2
\varepsilon_\mu{}^{ab} \pi^b \xi^c + 2m\alpha_1 B_\mu \xi, \qquad
\delta_{11} A_\mu = 2ma_3 \varphi \xi_\mu \label{c11} \\
\delta_{11} \pi^a &=& - 2ma_3 \varepsilon^{abc} B^b \xi^c + 3m\alpha_2
\pi^a \xi, \qquad \delta_{11} \varphi = m\alpha_2 \varphi \xi
\nonumber
\end{eqnarray}
Now let us turn to the variations of order $m^2$. Additional terms to
Lagrangian look as follows:
\begin{equation}
{\cal L}_{12} = c_1 \epthree f_\mu{}^a f_\nu{}^b f_\alpha{}^c + c_2
\eptwo f_\mu{}^a f_\nu{}^b \varphi + c_3 f \varphi^2 + c_4 \varphi^3
\label{l12}
\end{equation}
{\bf $\eta^a$ transformations:}
$$
\varepsilon^{\mu\nu\alpha} [ (6c_1 + 2m^2\alpha_1 - 2M^2\kappa_3 )
f_{\mu,\nu} f_\alpha{}^a \eta^a + (2c_2 - 4m^2a_3 - 4mM\kappa_3)
f_{\mu,\nu} \eta_\alpha \varphi ]
$$
This gives us:
$$
c_1 = \frac{(M^2+m^2)}{3}, \qquad c_2 = 2m^2a_3 + 2mM\kappa_3 = -
2mM\alpha_2
$$
{\bf $\xi^a$ transformations:}
\begin{eqnarray*}
&& - 6c_1 \epthree D_\mu f_\nu{}^a f_\alpha{}^b \xi^c + 6c_1
\varepsilon^{\mu\nu\alpha} [ \Omega_\mu{}^a \xi^a f_{\nu,\alpha} +
f_\mu{}^a \Omega_\nu{}^a \xi_\alpha] + \\
&& + 2c_2 \eptwo D_\mu f_\nu{}^a \xi^b \varphi - 2c_2
\varepsilon^{\mu\nu\alpha} \Omega_{\mu,\nu} \xi_\alpha \varphi - \\
&& - 2(c_2+mM\alpha_2) \eptwo f_\mu{}^a D_\nu \varphi \xi^b -
2mM\alpha_2 \eptwo f_\mu{}^a \xi_\nu \pi^b  - 2mM\alpha_2 \eptwo \pi^a
\xi^b f_{\mu,\nu}  - \\
&&- 2c_3 \xi^\mu \varphi D_\mu \varphi - 7m^2\alpha_2
\varphi (\pi\xi) + 4mMa_3 \varphi (\pi\xi) 
\end{eqnarray*}
Here we need the following corrections:
\begin{eqnarray}
\delta_{12} \Omega_\mu{}^a &=& - 6c_1 \varepsilon^{abc} f_\mu{}^b
\xi^c + 2c_2 \varepsilon_\mu{}^{ab} \varphi \xi^b \nonumber \\
\delta_{12} \pi^a &=& 2mM\alpha_2 (\xi^\mu f_\mu{}^a - f \xi^a) + 2c_3
\varphi \xi^a \label{c12}
\end{eqnarray}
Then all variations cancel provided
$$
2c_3 + 7m^2\alpha_2 - 2m^2 \kappa_3 = 0
$$
In this, all $\xi$ variations also cancel.

We still have variations of order $m^3$. As we have checked variations
for $\xi^a$ transformations cancel, while for $\xi$ transformations
we get
$$
6 (mc_3 - Mc_4 + m^3\alpha_2) \varphi^2 \xi
$$
This gives us an expression for last unknown coefficient $c_4$:
$$
Mc_4 = m^3 (\kappa_3 - \frac{5}{2} \alpha_2)
$$

Thus we have complete set of cubic vertices (\ref{l10}), (\ref{l11}),
(\ref{l12}) and corresponding corrections to gauge transformations
(\ref{c10}), (\ref{c11}), (\ref{c12}). Note that as we will see in the
next subsection for usual gravitational interactions we must have
$\alpha_1 = \alpha_2 = - 2\kappa_3$, but for massive spin 2
self-interaction we obtained\footnote{In agreement with the general
results from metric-like formalism \cite{Zin06}.}
$$
\alpha_1 = - \kappa_3, \qquad \alpha_2 = - (1 + \frac{m^2}{2M^2})
\kappa_3
$$
This result is a consequence of spontaneously broken symmetries with
Stueckelberg fields providing their non-linear realization. From the
last relation above it follows that it is impossible to take partially
massless limit $M \to 0$ in the interacting theory.

Contrary to the massless case due to the presence of spin 1 and spin 0
fields there exist (and necessarily must be present) quartic and
higher vertices so that the results obtained is not complete theory
yet. We will return to this point in Subsection 2.4.

\subsection{Gravitational interaction}

In this subsection we consider gravitational interactions for massive
spin 2 particles, i.e. cross-interaction for massless and massive
ones. We have explicitly checked that the only solution possible
exactly corresponds to standard minimal gravitational interactions,
i.e. can be obtained by the usual rule where background frame
$e_\mu{}^a$ is replaced by dynamical one $h_\mu{}^a$ while $AdS$
covariant derivatives are replaced by fully Lorentz covariant ones.
Thus we will not give details of calculations here (they are similar
to those in previous subsection) and just present the final results.
Here complete set of cubic vertices also consists of three parts:
\begin{equation}
{\cal L}_1 = {\cal L}_{10} + {\cal L}_{11} + {\cal L}_{12}
\end{equation}
\begin{eqnarray*}
{\cal L}_{10} &=& \kappa_2 \epthree [ h_\mu{}^a \Omega_\nu{}^b
\Omega_\alpha{}^c + 2 f_\mu{}^a \omega_\nu{}^b \Omega_\alpha{}^c ] +
\kappa_2 h B^a B^a - 2\kappa_2 \varepsilon^{\mu\nu\alpha} h_\mu{}^a
B^a D_\nu A_\alpha - \\
 && - \kappa_2 h \pi^a \pi^a + 2\kappa_2 \eptwo h_\mu{}^a D_\nu
\varphi \pi^b \\
{\cal L}_{11} &=& 2m\kappa_2 \varepsilon^{\mu\nu\alpha} [ 2 h_\mu{}^a
\Omega_\nu{}^a A_\alpha + h_\mu{}^a B^a f_{\nu,\alpha} - B_\mu
h_\nu{}^a f_\alpha{}^a ] + 4M\kappa_2 \eptwo h_\mu{}^a A_\nu \pi^b  \\
{\cal L}_{12} &=& M^2\kappa_2 \epthree h_\mu{}^a f_\nu{}^b
f_\alpha{}^c + 4mM\kappa_2 \eptwo h_\mu{}^a f_\nu{}^b \varphi +
6m^2\kappa_2 h \varphi^2
\end{eqnarray*}
while appropriate corrections to gauge transformations look as
follows:
\begin{equation}
\delta_1 = \delta_{10} + \delta_{11} + \delta_{12}
\end{equation}
\begin{eqnarray*}
\delta_{10} \omega_\mu{}^a &=& - 2\sigma\kappa_2 \varepsilon^{abc}
\Omega_\mu{}^b \eta^c, \qquad 
\delta_{10} h_\mu{}^a = - 2\sigma\kappa_2 \varepsilon^{abc} [
f_\mu{}^b \eta^c + \Omega_\mu{}^b \xi^c ] \\
\delta_{10} \Omega_\mu{}^a &=& - 2\kappa_2 \varepsilon^{abc} [
\Omega_\mu{}^b \hat{\eta}^c + \omega_\mu{}^b \eta^c ] \\
\delta_{10} f_\mu{}^a &=& - 2\kappa_2 \varepsilon^{abc} [ f_\mu{}^b
\hat{\eta}^c + \Omega_\mu{}^b \hat{\xi}^c + h_\mu{}^b \eta^c +
\omega_\mu{}^b \xi^c ] \\
\delta_{10} B^a &=& - 2\kappa_2 \varepsilon^{abc} B^b \hat{\eta}^c +
2\kappa_2 \hat{\xi}^b D^a B^b, \qquad
\delta_{10} A_\mu = 2\kappa_2 \varepsilon_\mu{}^{ab} B^a \hat{\xi}^b
\\
\delta_{10} \pi^a &=& - 2\kappa_2 \varepsilon^{abc} \pi^b \hat{\eta}^c
+ 2\kappa_2 (\hat{\xi}^\mu D_\mu \pi^a - \hat{\xi}^a (D\pi)), \qquad
\delta_{10} \varphi = 2\kappa_2 (\pi\hat{\xi})
\end{eqnarray*}
\begin{eqnarray*}
\delta_{11} \omega_\mu{}^a &=& 2m\sigma\kappa_2 ( 2A_\mu \eta^a - 
B_\mu \xi^a - 2 \Omega_\mu{}^a \xi) \\
\delta_{11} \Omega_\mu{}^a &=& 2m\kappa_2 (B_\mu \hat{\xi}^a -
e_\mu{}^a (B\hat{\xi})) \\
\delta_{11} f_\mu{}^a &=& 4m\kappa_2 ( - A_\mu \hat{\xi}^a + h_\mu{}^a
\xi) \\
\delta_{11} B_\mu &=& 4\kappa_2 ( m \Omega_\mu{}^a \hat{\xi}^a +
M \varepsilon_\mu{}^{ab} \pi^a \hat{\xi}^b) \\
\delta_{11} A_\mu &=& 2m\kappa_2 ( - f_\mu{}^a \hat{\xi}^a + h_\mu{}^a
\xi^a  )
\end{eqnarray*}
\begin{eqnarray*}
\delta_{12} \omega_\mu{}^a &=& 2M\sigma\kappa_2 ( - M
\varepsilon^{abc} f_\mu{}^b \xi^c + 2m\varepsilon_\mu{}^{ab} \varphi
\xi^b ) \\
\delta_{12} \Omega_\mu{}^a &=&  2M\kappa_2 ( - M \varepsilon^{abc}
f_\mu{}^b \hat{\xi}^c + 2m \varepsilon_\mu{}^{ab} \varphi
\hat{\xi}^b - M \varepsilon^{abc} h_\mu{}^b \xi^c ) \\
\delta_{12} \pi^a &=& 4m\kappa_2 [ - M (\hat{\xi}^\mu f_\mu{}^a - f
\hat{\xi}^a) + 3m \varphi \hat{\xi}^a ]
\end{eqnarray*}

Note that in this case nothing prevent us from taking partially
massless limit $M \to 0$ so that at least in the linear approximation
it is possible to obtain gravitational interactions for partially
massless spin 2 particles.

\subsection{Beyond linear approximation}

As we have already mentioned, due to the presence of spin 1 and spin 0
components there must be quartic (and even higher) vertices. So,
contrary to the massless case, the linear approximation considered in
the two previous subsections is not the end of the story. But all the
terms that includes spin 2 fields only have already been fixed. Thus,
if we require that the model we are looking for does admit
non-singular massless limit, we may try to put some restriction on the
parameters. Recall that in the massless case we have
$$
{\cal L}_{10} = \epthree [ \kappa_0 h_\mu{}^a \omega_\nu{}^b
\omega_\alpha{}^c + \kappa_2 h_\mu{}^a \Omega_\nu{}^b
\Omega_\alpha{}^c + 2\kappa_2 f_\mu{}^a \omega_\nu{}^b
\Omega_\alpha{}^c + \kappa_3 f_\mu{}^a \Omega_\nu{}^b
\Omega_\alpha{}^c ]
$$
and all quadratic variations for $\hat{\eta}^a$, $\hat{\xi}^a$,
$\eta^a$ and $\xi^a$ transformations cancelled provided $\kappa_2 =
\sigma\kappa_0$ with arbitrary $\kappa_3$. But in the massive case we
have additional symmetry:
\begin{eqnarray*}
\delta \omega_\mu{}^a &=& - 4m\sigma\kappa_2 \Omega_\mu{}^a \xi,
\qquad \delta \Omega_\mu{}^a = - 2m\kappa_3 \Omega_\mu{}^a \xi  \\
\delta f_\mu{}^a &=& 4m\kappa_2 h_\mu{}^a \xi + 2m\kappa_3 f_\mu{}^a
\xi
\end{eqnarray*}
where we collected all terms form both previous subsections. It was
not evident from the very beginning but it turns out that
cancellation for quadratic $\xi$ variations is indeed possible
provided the following relation holds:
$$
4\sigma\kappa_0{}^2 + \kappa_3{}^2 = 0 \quad \Longrightarrow \quad
\sigma = - 1
$$
As can easily be seen this one relation gives us two important
results. First, we get a relation between two previously independent
coupling constants. Second, this solution exists for $\sigma = - 1$
only when massless graviton is a ghost exactly as in the so  called
"New massive gravity".

\section*{Conclusion}

Thus we have seen that constructive approach based on the frame-like
gauge invariant description of massive spin 2 particles does allow
one to systematically investigate possible consistent ghost-free
models though due to large number of fields involved this requires
much more work than in the massless case. It is  evident that such
approach admits straightforward generalization to higher spins. In
this in three dimensional case we will have to nice features making
investigations simpler --- there no any quartic vertices for any spins
$s \ge 2$ and also there are no so called extra fields and thus there
is no need in higher derivatives. So we may hope to gain some useful
experience for work with massive higher spin fields. Also it is worth
noting that such approach can be applied to higher dimensional
theories as well, in this, first of all it would be interesting to
understand peculiar features of massive gravity and bigravity in
$d=4$. Work is in progress in both directions.

\vskip 1cm \noindent
{\bf Acknowledgment} 
The work was supported in parts by RFBR grant No.11-02-00814.

\appendix

\section{On cubic vertex with two spin 2 and one spin 0}

The most general form for such cubic vertex with two derivatives:
\begin{eqnarray*}
{\cal L}_1 &=& a_1 \eptwo \omega_\mu{}^a \omega_\nu{}^b \varphi + a_2
\epthree h_\mu{}^a h_\nu{}^b D_\alpha \pi^c + \\
 && + \varepsilon^{\mu\nu\alpha} [ a_3 \omega_\mu{}^a h_\nu{}^a
\pi_\alpha + a_4 \omega_{\mu,\nu} h_\alpha{}^a \pi^a + a_5
\omega_\alpha{}^a \pi^a h_{\nu,\alpha} + a_6 \omega_\mu{}^a D_\nu
h_\alpha{}^a \varphi + a_7 \omega_\mu{}^a h_\nu{}^a D_\alpha \varphi ]
\end{eqnarray*}
In this, there exist three possible field redefinitions:
$$
\omega_\mu{}^a \Rightarrow \omega_\mu{}^a + \rho_1 \varphi
\omega_\mu{}^a + \rho_2 \varepsilon^{abc} h_\mu{}^b \pi^c, \qquad
h_\mu{}^a \Rightarrow h_\mu{}^a + \rho_3 \varphi h_\mu{}^a
$$
Let us consider variations under $\hat{\eta}^a$ transformations:
\begin{eqnarray*}
\delta_{\hat{\eta}} {\cal L}_1 &=& (2a_1+a_6) \eptwo D_\mu
\omega_\nu{}^a \hat{\eta}^b \varphi + \\
 && + \varepsilon^{\mu\nu\alpha} [ - a_3 \hat{\eta}^a D_\mu h_\nu{}^a
\pi_\alpha + a_4 \hat{\eta}_\mu D_\nu h_\alpha{}^a \pi^a - a_5
(\pi\hat{\eta}) D_\mu h_{\nu,\alpha} - (a_6+a_7) \hat{\eta}^a D_\mu
h_\nu{}^a D_\alpha \varphi ] + \\
 && + \varepsilon^{\mu\nu\alpha} [ (2a_2+a_3) h_\mu{}^a \hat{\eta}^a
D_\nu \pi_\alpha + (2a_2-a_5) h_{\mu,\nu} \hat{\eta}^a D_\alpha \pi^a
- a_4 \hat{\eta}_\mu h_\nu{}^a D_\alpha \pi^a ] + \\
 && + \eptwo [ - a_3 \omega_\mu{}^a \pi_\nu \hat{\eta}^b + a_4
\omega_{\mu,\nu} \pi^a \hat{\eta}^b ] + a_5 \hat{\eta}^\mu
\omega_\mu{}^a \pi^a - (2a_1+a_6+a_7) \eptwo \omega_\mu{}^a D_\nu
\varphi \hat{\eta}^b
\end{eqnarray*}
From the third line it follows that $a_3 = - 2a_2$, $a_4 = 0$, $a_5 =
2a_2$. But in this case terms with coefficients $a_{2,3,5}$ can be
removed by redefinition with parameter $\rho_2$. This leaves us with
$$
(2a_1+a_6) \eptwo D_\mu \omega_\nu{}^a \hat{\eta}^b \varphi -
(a_6+a_7) \varepsilon^{\mu\nu\alpha} D_\mu f_\nu{}^a \hat{\eta}^a
D_\alpha \varphi - (2a_1+a_6+a_7) \eptwo \omega_\mu{}^a \hat{\eta}^b
D_\nu \varphi
$$
As usual to compensate them we introduce corrections to gauge
transformations:
$$
\delta h_\mu{}^a = \alpha_1 \varepsilon_\mu{}^{ab} \hat{\eta}^b
\varphi, \qquad \delta \omega_\mu{}^a = \alpha_2 D_\mu \varphi
\hat{\eta}^a
$$
They give additional contribution:
$$
- \alpha_1 \eptwo D_\mu \omega_\nu{}^a \hat{\eta}^b \varphi -
\alpha_2 \varepsilon^{\mu\nu\alpha} D_\mu f_\nu{}^a \hat{\eta}^a
D_\alpha \varphi + \alpha_2 \eptwo \omega_\mu{}^a \hat{\eta}^b
D_\nu \varphi
$$
Hence $a_1+a_6+a_7 = 0$ and all remaining terms can be removed by
field redefinitions with parameters $\rho_1$ and $\rho_3$.

\section{On cubic vertex with two massless spin 2 and one massive one}

As it has been explained in the Subsection 2.2 we will look for cubic
vertices and appropriate corrections to gauge transformations in the
form
$$
{\cal L}_1 = {\cal L}_{10} + {\cal L}_{11} + {\cal L}_{12}, \qquad
\delta_1 = \delta_{10} + \delta_{11} + \delta_{12}
$$
Moreover, it turns out that to see that such vertex does not exist it
is enough to consider gauge transformations for massive spin 2 field
only. Taking into account results of Appendix A, the most general
possibility for ${\cal L}_{10}$ is:
$$
{\cal L}_1 = \kappa_1 \epthree [ f_\mu{}^a \omega_\nu{}^b
\omega_\alpha{}^c + 2 h_\mu{}^a \omega_\nu{}^b \Omega_\alpha{}^c ]
$$
while corresponding corrections to gauge transformations were given by
formula (\ref{ckap1}) in Subsection 1.2.

Let us turn to variations of order $m$. The most general additional
terms for the Lagrangian have the form:
$$
{\cal L}_{11} = \varepsilon^{\mu\nu\alpha} [ b_1 h_\mu{}^a
\omega_\nu{}^a A_\alpha + b_2 h_\mu{}^a B^a h_{\nu,\alpha} ]
$$
In this order there are no variations for $\eta^a$ and $\xi^a$
transformations while for $\xi$ transformations we obtain:
$$
\varepsilon^{\mu\nu\alpha} [ - b_1 D_\mu h_\nu{}^a \omega_\alpha{}^a
\xi + b_1 h_\mu{}^a D_\nu \omega_\alpha{}^a \xi ] + 2ma_1 \eptwo
\omega_\mu{}^a \omega_\nu{}^b \xi
$$
This variations can be compensated by the following corrections:
$$
\delta_1 \omega_\mu{}^a = - b_1 \omega_\mu{}^a \xi, \qquad
\delta_1 h_\mu{}^a = b_1 h_\mu{}^a \xi
$$
provided $b_1 = 2ma_1$. 

We proceed with the variation of order $m^2$ and introduce the last
part of the Lagrangian
$$
{\cal L}_{12} = c_1 \epthree h_\mu{}^a h_\nu{}^b f_\alpha{}^c + c_2
\eptwo h_\mu{}^a h_\nu{}^b \varphi
$$
Variations under $\xi^a$ transformations 
\begin{eqnarray*}
&& - 2c_1 \epthree D_\mu h_\nu{}^a h_\alpha{}^b \xi^c + 2mb_1
\varepsilon^{\mu\nu\alpha} h_\mu{}^a \omega_\nu{}^a \xi_\alpha + \\
&& + 2\kappa_1M^2 \varepsilon^{\mu\nu\alpha} [ h_\mu{}^a 
\omega_{\nu,\alpha} + h_{\mu,\nu} \omega_\alpha{}^a ] \xi^a -
2\kappa_1\lambda^2 \varepsilon^{\mu\nu\alpha} [ h_\mu{}^a 
\omega_{\nu,\alpha} \xi^a - h_\mu{}^a \omega_\nu{}^a \xi_\alpha ]
\end{eqnarray*}
require corrections
$$
\delta \omega_\mu{}^a = - 2c_1 \varepsilon^{abc} h_\mu{}^b \xi^c
$$
and we obtain:
$$
\varepsilon^{\mu\nu\alpha} [ \kappa_2(M^2-\lambda^2) h_\mu{}^a
\omega_{\nu,\alpha} \xi^a + (\kappa_2M^2-c_1) h_{\mu,\nu}
\omega_\alpha{}^a \xi^a + (\kappa_2m^2+\kappa_2\lambda^2 - c_1)
h_\mu{}^a \omega_\nu{}^a \xi_\alpha = 0
$$
It is easy to see that solution is possible for $m=0$ only.

\end{document}